\documentclass[12pt]{article}
 \usepackage{graphicx,psfrag,epsfig}

\makeatletter
\def\ifempty#1{\@ifempty #1\@emptymarkA\@emptymarkB}%
\def\@ifempty#1#2\@emptymarkB{\ifx #1\emptymarkA}%
\def\@emptymarkA{\@emptymarkA}%

\renewcommand{\part}[1]{%
    \stepcounter{part}%
    \begin{center}%
    {\large\bfseries\thepart.\ #1}
    \end{center}}%

\renewcommand{\@seccntformat}[1]{%
    {\csname the#1\endcsname}. \ 
    }

\renewcommand{\section}{%
    \@startsection{section}{1}{\z@}%
    {-3.5ex plus -1ex minus -.2ex}%
    {2.3ex plus.2ex}%
    {\centering\large\bfseries}}

\renewcommand{\subsection}{\@startsection{subsection}{2}{0pt}%
    {-3.25ex plus -1ex minus -.2ex}%
    {1.5ex plus .2ex}%
    {\centering\normalsize\bfseries}}

\renewcommand{\subsubsection}{\@startsection{subsubsection}{2}{0pt}%
    {-3.25ex plus -1ex minus -.2ex}%
    {1.5ex plus .2ex}%
    {\centering\normalsize\itshape}}

\renewenvironment{thebibliography}[1]
     {\part{REFERENCES CITED}%
      \list{\@biblabel{\@arabic\c@enumiv}}%
           {\settowidth\labelwidth{\@biblabel{#1}}%
            \leftmargin\labelwidth
            \advance\leftmargin\labelsep
            \@openbib@code
            \usecounter{enumiv}%
            \let\p@enumiv\@empty
            \renewcommand\theenumiv{\@arabic\c@enumiv}}%
      \sloppy\clubpenalty4000\widowpenalty4000%
      \sfcode`\.\@m}
     {\def\@noitemerr
       {\@latex@warning{Empty `thebibliography' environment}}%
      \endlist}

\@addtoreset{page}{chapter}
\def\@stpelt#1{\global\csname c@#1\endcsname%
    \expandafter\ifx \csname#1\endcsname \page%
        \@ne%
    \else%
        \z@ \fi}

\makeatother

\begin{document}
\begin{center}
{\Large\bf From Old Symmetries to New Symmetries:\\ Quarks, Leptons and B-L}

\begin{center}{
 {\bf Rabindra N. Mohapatra\footnote{Invited review to appear in the volume ``{\it Fifty years of Quarks}'' edited by H. Fritzsch  (to be published by World Scientific)}}}

{\it Maryland
Center for Fundamental Physics and Department of Physics,
University of Maryland, College Park, Maryland 20742, USA}
\end{center}
\date{\today}
\begin{abstract}
The Baryon-Lepton difference ($B-L$) is increasingly emerging as a possible new symmetry of the weak interactions of quarks and leptons as a way to understand the small neutrino masses. There is the possibility that current and future searches at colliders and in low energy rare processes may provide evidence for this symmetry. This paper provides a brief overview  of the early developments that led to  B-L  as a possible symmetry beyond the standard model, and also discusses some recent developments.
\end{abstract}
\end{center}


\section{Early history}  Progress in physics comes in many ways. Sometimes theories follow experiments and sometimes it comes the other way. Classic examples of the first type   are, e.g. Faraday's law and Oersted's discovery of connection between electricity and magnetism to name but two. There are also equally illustrious example of experiments following theory : Hertz's discovery of electromagnetic waves following the suggestion of Maxwell, and a more recent example of neutrino being discovered almost 25 years after the suggestion made by Pauli. The same pattern of close entanglement between theory and experiment, with one influencing the other, has continued in the 20th and onto the current century. Theoretical insights into physical phenomena have often followed from the application of novel mathematical techniques e.g differential equations, algebras, group theory, to name a few. Again in this area too, mathematical developments have followed from physics and vice versa (consider for example the development of calculus by Newton as a way to describe motion).

In the second half of the 20th century, group theory has played an important role in the development of physics as new symmetries were discovered in a variety of physical systems. These  provided the primary guiding light for fundamental areas of physics such as elementary particles and condensed matter physics and even the posteriori understanding of some results in quantum mechanics, e.g. the role of $O(4)$ symmetry in the hydrogen atom. In the domain of elementary particle physics, the discovery of the quark model of hadrons and of the standard model of electroweak interactions are examples where symmetries played a triumphant role. This success strengthened the belief that there may be newer symmetries in nature that will be manifested  as we move to uncover physics at ever smaller distances. Many attempts were made to combine space-time symmetries with internal symmetries, e.g., theories based on $SU(6)$ and in the 1970s the emergence of supersymmetry, which from 1980's  became the dominant theme in both theory and experiment. Although experiments ultimately decide which symmetries live and which die, either way they leave a lasting impact on the field. In this article, I will focus on  a new symmetry of particle physics, the $B-L$ symmetry, which is a global symmetry of the standard model and appears to be emerging as a local symmetry designed for understanding the physics of neutrino masses. 

This article is organized as follows. In sec. 1.2 , I review the history of how symmetries started to enter particle and nuclear physics, how they slowly determined the subsequent developments in the field and how $B-L$ started to make its appearance from the apparent similarity between hadrons and leptons. Sec. 1.3 briefly discusses the standard model (SM) of electroweak interactions and the clear indication of a new $B-L$ symmetry in nature. Section 1.4, discusses the suggestion that $B-L$ plays the role of a gauge symmetry, once the right handed neutrinos are introduced into the standard model, a realization that followed only after the left-right symmetric models of weak interactions were introduced making the existence of the right handed neutrino automatic and its defining role in making $B-L$ gauge symmetry theoretically consistent.  This preceded by several years the developments in understanding of its role in neutrino mass. Sec. 1.5 discusses the connection between small neutrino masses and the breaking of $B- L$ symmetry, followed in sec 1.6, by the prediction of a new baryon number violating process once the left-right symmetric models are embedded into a quark-lepton unified version of the model. Sec. 1.7, is devoted to a model-independent connection between proton decay, neutron oscillation and Majorana neutrinos, where $B-L$ breaking plays an important role. In the concluding section, we note briefly searches for the $B-L$ symmetry in various experiments and future prospects for success of such searches.

\section{Old symmetries and quark-lepton similarity}
The field of particle physics was born in the 1950s as more and more particles beyond the familiar neutron, proton, and the $\pi$ meson were discovered in cyclotrons and in cosmic rays. They included  both neutral and charged K-mesons (discovered in 1947), the hyperon $\Lambda$ (discovered in 1950) and $\Sigma^{\pm,0}$  and $\Xi$ (both in 1953). The $\rho$ meson was discovered in 1961 following theoretical suggestions  (an early example of experiment following theory). This was followed shortly thereafter by $\phi$, $K^*$ and the $\omega$ vector mesons. As the number of new particles kept increasing, there clearly was  need to understand their fundamental nature and systematize their study and possibly predict more new particles from such studies. That is precisely what happened in the 1960s when, following the iso-spin symmetry suggested by Heisenberg, Gell-Mann and Ne'eman proposed the SU(3) symmetry of strong interactions as a way for classifying the new particles and studying their properties.
While isospin was based on the internal symmetry $SU(2)$, which covered only the particles $p,n,\pi^{\pm,0}$, the goal of $SU(3)$ was more ambitious; it was supposed to explain many of the newly discovered mesons and baryons in terms of irreducible representations of an internal $SU(3)$ symmetry and in that process understand their masses and decay properties. The Gell-Mann Okubo mass formula introduced to understand the masses was a phenomenal success, and predicted the $\Omega^-$ particle which was discovered in 1964 at the Brookhaven National Laboratory, providing thereby a striking confirmation of the relevance of the symmetry approach to particle physics.
This symmetry approach ultimately led to the suggestion  by Gell-Mann and Zweig~\cite{gz} in 1964 that the fundamental building blocks of all hadrons (e.g., $p, n, \pi,...$) are tinier particles called quarks. 

As developments were taking place in hadron physics, a quiet revolution was taking shape in the domain of leptons i.e., electrons, neutrinos ($\nu_e$) etc. The electron neutrino, proposed by Pauli in 1930, was discovered in 1957 by Reines and Cowan. The muon, a close but heavier cousin of the electron was discovered in cosmic rays in 1947. It was realized already by Pauli and Fermi that in nuclear $\beta$ decay an electron is accompanied by the antineutrino ($\bar{\nu}_e$). Also bombarding the nucleus by this produced neutrino state only produced positrons and not electrons suggesting that there was a difference between neutrino and its antiparticle $\bar{\nu}_e$ produced produced in nuclear $\beta$ decay.
Similar situations had been encountered before with the proton in that the hydrogen atom was stable but there was no apparent reason for it to be that way. This led Stuckelberg to propose in 1938 that there must be a new conserved quantum number, the baryon number (denoted by $B$). This meant that in any physical process, both the initial and the final states must have the same baryon number. This also keeps proton as a stable particle, in agreement with observations. Subsequent discovery of the other baryons, e.g., $\Lambda$, $\Sigma$, etc which decay only transform to other baryons, e.g. protons and neutrons added richness to the concept of the baryon number. The fundamental origin of this quantum number has been one of the mysteries of theoretical physics for a very long time. 

The fact that the electron in the hydrogen atom remains stable can also be understood in a similar manner by postulating another quantum number (called lepton number $L$). Of course if only the electron carried a lepton number, it would not have been very interesting but, as noted above, the antineutrino, which is emitted simultaneously with the electron in $\beta$ decay, also seemed to have this property that, in its subsequent scattering from nuclei produces only a positron and not an electron. This could be understood easily if the neutrino also carried the same lepton number as the electron, with the positron and the $\bar{\nu}_e$ carrying a negative lepton number. This way, nuclear $\beta$ decay, where $n\to p+e^-+\bar{\nu}_e$, both baryon number and lepton numbers are conserved. Thus far, we have not assigned any particular value to $B$ and $L$ for the above particles, and, without any loss of generality, we can assign $B=1$ to $p, n, \Lambda, \Sigma...$ etc and $L=+1$ to $\nu_e, e,....$. According to the quark model, three quarks would form a baryon such as the proton implying that a quark would carry $B=\frac{1}{3}$. No process involving elementary particles has been observed to violate either baryon or lepton number conservation. There are nevertheless, circumstantial indications,  that these laws must be broken. Most compelling of them is the fact the universe seems to have an excess of baryons (matter) over anti-baryons (anti-matter) and the theoretical recognition that non-perturbative effects in the standard model can lead to the violation of both baryon and lepton number.

As $SU(3)$ symmetry of hadrons was gaining a firm hold in the physics of baryons and mesons, a question was being raised regarding whether there were any similar symmetries in the domain of leptons. However unlike the multiplicity of baryons and mesons, until the beginning of 1960s, only three leptons were known to exist: $e^-, \mu^-$ and $\nu_e$ and their anti-particles and no more. Clearly, there was no need for a symmetry like $SU(3)$. A curious feature none-the-less was noted in 1959 by Gamba, Marshak and Okubo~\cite{gmo} that the three baryons $(p, n, \Lambda)$ were arranged in electric charge (and very crudely in mass) roughly the same way as the leptons $(\nu_e, e^-, \mu^-)$. This led them to suggest that there was a symmetry between baryons and leptons (modulo an overall shift in the charge values). They called this baryon lepton symmetry. The shifted charge pattern could be understood by writing the following formula:
\begin{eqnarray}\label{BLQ}
Q~=~I_3+(T+B-L)/2
\end{eqnarray}
where $(p,n)$ and $(\nu_e, e)$ are assigned to some new isospin group, with $I_3$ being the third isospin generator;  the $\mu^-$ and $\Lambda$ in this approach are considered 
iso-singlets. The quantum number $T$ is like strangeness with $T=-1$ for both $\Lambda$ and $\mu^-$~\cite{REM}. Equation (1.1)  is a unified formula for both hadrons and leptons. Once the quark model was introduced,  quark lepton symmetry  could be used instead of baryon lepton symmetry and one would write $(u,d, s)$ instead of $(p, n, \Lambda)$ and the same electric charge formula would apply. Quark-lepton symmetry then became a reflection of the symmetry
\begin{eqnarray}
\left(\begin{array}{c}u\\d\\s\\\end{array}\right)\leftrightarrow\left(\begin{array}{c}\nu_e\\e\\\mu\\\end{array}\right)
\end{eqnarray}

 A beautiful aspect of quark lepton symmetry is that it connects two kinds of elementary particles and suggests that there is a separate weak $SU(2)$ symmetry (to be contrasted with the the familiar one operating on protons, neutrons, pions etc ) that also operates on leptons. For the first time, a symmetry appeared in the lepton sector. A new symmetry $SU(2)_W$ was born from the old hadronic $SU(2)$ of strong interactions. Eventually, this new $SU(2)_W$ was identified as a local symmetry and became the corner stone of the standard model of Glashow, Weinberg and Salam~\cite{gws}. 

A major transformation came over this formulation of quark-lepton symmetry when in 1962, Lederman et al discovered the muon neutrino $\nu_{\mu}$. It appeared distinct from the $\nu_e$, and therefore the three quarks needed another partner for quark-lepton symmetry still was to work. This led Bjorken and Glashow~\cite{bg} in 1964 to postulate that there must be a heavier up-like quark which now a days we know to be the charm quark. Discovery of charm quark at SLAC and Brookhaven maintained the quark-lepton symmetry in its intended form. Clearly, the formula for electric charge had to be rewritten. Is $B-L$ going to remain as part of the formula ? The answer had to be postponed. Meanwhile, these and other developments were slowly ushering in a new era in particle physics, which not only determined the spectacular developments that dominated the field for the next fifty years, but are also likely to continue their impact in the future.

\section{$B-L$ symmetry and the standard model}
 Glashow, Weinberg and Salam\cite{gws} recognized that this new $SU(2)_W$ (or $SU(2)_L$) symmetry is a local symmetry whose associated gauge bosons can mediate the weak interactions. They proposed a gauge theory based on the $SU(2)_L\times U(1)_Y$ group that at the lowest tree level had the right properties to describe the known $V-A$ form for the charged current weak interactions; the extra $U(1)_Y$ group was needed to unify electromagnetism with weak interactions.  We now recognize this theory as the standard model of weak and electromagnetic interactions. This model has been confirmed by experiments, the latest being the discovery of the Higgs boson at the LHC. Below we give a brief overview of some of the symmetry aspects of the model. Under the weak $SU(2)_L\times U(1)_Y$ group, the fermions of one generation are assigned as follows:
\begin{eqnarray}
Q_L~= \left(\begin{array}{c} u_L\\d_L \end{array}\right) \equiv (1/2, 1/3); ~~~L~= \left(\begin{array}{c} \nu_L\\e_L \end{array}\right)\equiv (1/2, -1);\\ \nonumber u_R\equiv (1,4/3); ~~~d_R\equiv (1,-2/3); ~~~e_R\equiv (1,-2)
\end{eqnarray}
where $u, d, \nu, e$ are the up and down quarks and the neutrino and electron fields, respectively. The subscripts $L, R$ stand for the left and right handed spin  chiralities of the corresponding fermion fields. The numbers in the brackets denote the $SU(2)_L$ and $Y$ quantum numbers.
There are four gauge bosons $W^{\pm}_\mu, W^3_\mu$,and $ B_\mu$ associated with the four generators of the gauge group, The interactions of these gauge fields with matter (the quarks and leptons) are determined by the symmetry of the theory and lead to the current-current form for weak interactions via exchange of the $W^\pm$ gauge boson. Before symmetry breaking, the gauge bosons and fermions are massless. The masslessness of the gauge bosons follows in a way similar to that of photon being massless in QED. Since  fermion mass terms correspond to bilinears of the form $\bar{\psi}_L\psi_R$ that connect the left and right chirality states of the fermion, such terms are forbidden by gauge invariance since the left chiral states of fermions in SM are $SU(2)_L$ doublets whereas the right chirality ones are singlets. 

To give them mass, we adopt the model for spontaneous breaking of gauge symmetry through the inclusion of  scalar fields~\cite{BEHGHK}, $\phi(1/2, 1)$ in the theory which transform as doublets of the gauge group (or weak isospin 1/2). This allows Yukawa couplings of the form $\bar{Q}\phi d_R$, $\bar{\psi}_L\phi e_R$ and $\bar{Q}\tilde{\phi} u_R$ where $Q$ refers to doublets of quarks and $\psi$ refers to doublets of leptons, and $\tilde{\phi}=i\tau_2\phi^*$ ($\tau_{1,2,3}$ denote the three Pauli matrices). If the gauge symmetry is broken by assigning a ground state value of the field $\phi$ as $<\phi >=\left(\begin{array}{c}0\\ v\end{array}\right)$, this gives mass not only to all the fermions in the theory but also to the gauge bosons $W^\pm$ and to $Z\equiv \cos\theta_W W_3+\sin\theta_W B$ where $\theta_W=tan^{-1}\frac{g^\prime}{g}$ is the weak mixing angle. It should be noted that  $<\phi>$ leaves one gauge degree of freedom unbroken i.e. 
\begin{eqnarray}\label{SMQ}
Q=I_3+\frac{Y}{2}
\end{eqnarray} 
(where $I_3$ denotes the third weak isospin generator of the $SU(2)_L$ gauge group).  since $Q<\phi>=0$, $Q$ remains an unbroken symmetry and  can be identified as the electric charge. Given the quantum numbers assigned to different particles, it reproduces the observed electric charges of all the particles of SM.  There is however a very unsatisfactory aspect  to this electric charge formula due to the presence of the ``fluttering'' $Y$ term which can be assigned at random. As a result, origin of the electric charge remains mysterious in the SM.

A point worth noting is that to make SM consistent with low energy observations such as extreme suppression of flavor changing neutral current (FCNC) process $K^0_L\to \mu^+\mu^-$, the charm quark had to be invoked. This is the celebrated Glashow-Illiopoulos-Maiani (GIM) mechanism~\cite{GIM} for suppressing the FCNC, which provided the true dynamical role of the charm quark beyond just, being the quark partner of the $\nu_{\mu}$ as noted above.

A major expectation of the standard model was that it should be renormalizable so that as in Quantum electrodynamics, it can make testable predictions. Chiral gauge theories however were known from late 1960s to have the Adler-Bell-Jackiw anomalies~\cite{adler}. The existence of such anomalies implies that an apparently conserved current at the tree level is no more conserved once quantum effects are taken into account and would thereby undo the renormalizability of the theory. The model as
constructed above turns out to be free of these gauge anomalies. Asking whether there are any other symmetries that are exact, brings to mind two natural candidates of great interest: the baryon ($B$)and lepton ($L$) numbers. Clearly, the tree level Lagrangian respects these symmetries separately; however at the quantum level (one loop), there are triangle diagrams that make both the B and L  currents anomalous i.e. for one generation of fermions, keeping only the $SU(2)_L$ contributions, we have
\begin{eqnarray}
\partial^\mu J_{B,\mu}=\partial^\mu J_{L,\mu}=\frac{g^2}{32\pi^2}W^{\alpha\beta}\tilde{W}_{\alpha\beta}
\end{eqnarray}
which means that they are separately not conserved but interestingly, the combination $B-L$ is conserved i.e.
\begin{eqnarray}
\partial^\mu \left(J_{B,\mu}- J_{L,\mu}\right)=0
\end{eqnarray}
The freedom from anomalies implies that no non-perturbative effect can break this symmetry.Thus $B-L$ symmetry is back in play as a symmetry for quark lepton physics and from a completely different framework. In the standard model, $B-L$ is not free of cubic i.e. anomalies in $Tr(B-L)^3\neq 0$, when summed over all fermions in the SM. Rather it is only free of linear anomalies i.e. $Tr(U(1)_{B-L}[SU(2)_L]^2)=0$ and $Tr(U(1)_{B-L}[U(1)_Y]^2)=0$. This means that $B-L$ is not a hidden local symmetry of standard model but rather just an exact global symmetry as noted.

One important consequence of $B-L$ being an exact global symmetry is that the non-perturbative effects in the SM known as sphaleron effects do break $B$ and $L$ separately but not $B-L$, as pointed out by 't Hooft ~\cite{th} . They can be represented as a twelve fermion gauge invariant operator $QQQQQQQQQLLL$ which breaks both baryon and lepton number but conserves $B-L$. The strength of this interaction is very weak at zero temperature but is much stronger in the early universe and has important implications for cosmology.

\section{B-L as the $U(1)$ gauge generator of weak interactions and new electric charge formula}
A key prediction of the standard model (SM) is that neutrino masses vanish since, unlike other fermions, which have both left and right handed chiralities in the theory, there is no right handed neutrino but just the left handed $SU(2)_L$ partner of $e_L$. Because several experiments have confirmed since 1990's that neutrinos have mass, the simplest extension of SM is to add to it one right handed neutrino per generation to account for this fact. As soon as this is done, one not only has $Tr(U(1)_{B-L}[SU(2)_L]^2)=0$ and $Tr(U(1)_{B-L}[U(1)_Y]^2)=0$ but also $Tr(B-L)^3= 0$. This allows for the possibility of gauging the $U(1)_{B-L}$ quantum number, which gives $B-L$ a dynamical role. 

Addition of a right handed neutrino suggests that weak interaction theory is not only quark-lepton symmetric but can also be written in a way that it conserves parity. These are the left-right symmetric models (LRS) which were written down in 1974-75~\cite{mp}.
The gauge group of the LRS model is: $SU(2)_L\times SU(2)_R\times U(1)_{B-L}$ which includes discrete parity symmetry and fermion assignments given by:
\begin{eqnarray}
Q_{L,R}~= \left(\begin{array}{c} u\\d \end{array}\right)_{L,R} (1/2,0, 1/3)~or~ (0,1/2,1/3);\\ \nonumber \psi_{L,R}~= \left(\begin{array}{c} \nu\\e \end{array}\right)_{L,R}\equiv (1/2,0, -1);~or~ ((0,1/2,-1)
\end{eqnarray}
That is because, under parity inversion left-handed fermions go to right handed fermions, the above assignment is parity symmetric. The resulting weak interaction Lagrangian is given by:
\begin{eqnarray}
{\cal L}_{wk}~=~i\frac{g}{2}\left(\bar{Q}_L\vec{\tau}.\vec{W}^\mu_L\gamma_\mu Q_L+\bar{\psi}_L\vec{\tau}.\vec{W}^\mu_L\gamma_\mu \psi_L\right)~+~L\leftrightarrow R
\end{eqnarray}
Clearly under parity inversion, if we transform $W_L\to W_R$, the Lagrangian is parity conserving. However once symmetry breaking is turned on,  $W_R$ will acquire a higher mass and introduce parity violation into low energy weak interaction. The effective weak interaction Hamiltonian below the $W$ boson mass can be written as:
\begin{eqnarray}
{\cal H}_I~=~\frac{g^2}{2M^2_{W_L}}({\cal J}^{+,\mu}_L{\cal J}^{-}_{\mu,L})+\frac{g^2}{2M^2_{W_R}}({\cal J}^{+,\mu}_R{\cal J}^{-}_{\mu,R})~+~h.c.
\end{eqnarray}
Note that if $m_{W_R}\gg m_{W_L}$, the above weak interactions violate parity almost maximally since the right handed current effects are suppressed by a factor $\frac{m^2_{W_L}}{m^2_{W_R}}$. This is a fundamentally different approach to observed parity violation than the one espoused in the SM.

In fact it was pointed out independently by Marshak and me ~\cite{MM} and Davidson~\cite{david}, that the electric charge formula now becomes
\begin{eqnarray}
Q~=~I_{3L}+I_{3R}+\frac{B-L}{2}
\end{eqnarray}
This is a considerable improvement over the SM electric charge formula of Eq.\ref{SMQ} in the sense that all terms in the formula are determined through physical considerations of weak, left and right isospin, and baryon and lepton number, that reflect independent characteristics of the various elementary particles. No freely floating parameters are needed to fix electric charges as in the standard model. Electric charge is no more a free parameter but is connected to other physical quantum numbers in the theory. As a result,  a number of interesting implications follow. I discuss them below.

\section{Neutrino mass and B-L symmetry}
Discovery of neutrino oscillations has confirmed that neutrinos have mass requiring therefore an extension of the standard model. If we simply add a $\nu_R$ and construct the usual Dirac mass for the neutrino in the same way as for the other fermions in SM, the  accompanying Yukawa coupling $h_\nu$  has to be order $10^{-12}$ to match observations. Perhaps this suggests that there is some new physics even beyond adding the $\nu_R$ to SM that will not require such small parameters. This is where the seesaw mechanism~enters~\cite{seesaw}, which  demands that the $\nu_R$'s have a large Majorana mass. Since the neutrinos have no electric charge, a Majorana mass for them is compatible with electric charge conservation, which seems to be an absolute symmetry of nature.  Once the this is done, the $\nu_L-\nu_R$ mass matrix is given by:
\begin{eqnarray}  
{\cal M}_{\nu, N}~=~\left(\begin{array}{cc} 0 & m_D\\ m^T_D & M_R\end{array}\right)
\end{eqnarray}
where each of the entries are $3\times 3$ matrices corresponding to three generations of fermions observed in nature. The Feynman diagram responsible for this is given in  figure 1:
\begin{figure}
\centerline{\includegraphics[width=4.5cm]{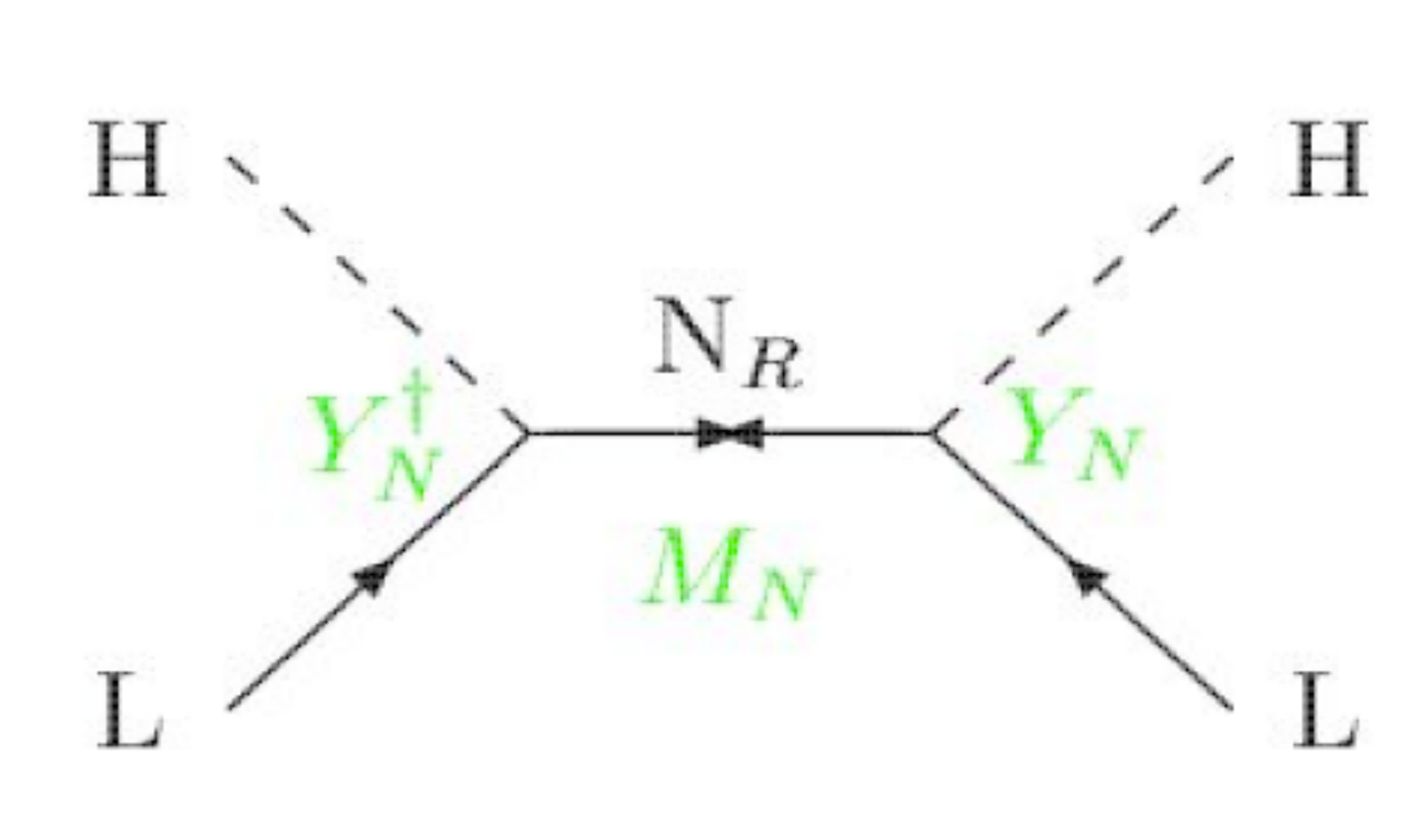}}
\caption{ Feynman diagram for the seesaw mechanism for understanding small neutrino masses. $Y_N$ are the Yukawa couplings of the right handed neutrino $N_R$  } \label{ra_fig1}
\end{figure}

Diagonalization of this mass matrix leads to a mass formula for the light neutrinos of the form
\begin{eqnarray}
{\cal M}_\nu~=~-m^T_D M^{-1}_R m_D
\end{eqnarray}
As $M_R$ corresponds to the right handed neutrino mass, it is not restricted by physics of the SM and can be large whereas $m_D$ is proportional to the scale of standard electroweak symmetry breaking and therefore of the same order as the quark-lepton masses. Thus by making $M_R$ large, we can obtain a very tiny neutrino mass. 

This raises two questions: (i) whether it is  a bit adhoc to add the right handed neutrinos ? and (ii) is there a physical origin of the seesaw scale or we just accept it as an arbitrary input into the theory ? We see below that both these questions are answered in the left-right symmetric extension of the SM, introduced for the purpose of understanding origin of parity violation in Nature.

The left-right model provides a reason why the $\nu_R$ should exist and the seesaw scale is given by the scale of parity breaking. In terms of the electric charge formula, we see that since $\Delta Q=0$ and above the SM scale, $\Delta I_{3L}=0$, we get from Eq. (1.12)
\begin{eqnarray}\label{BL}
\Delta I_{3R}\simeq -\Delta\left(\frac{B-L}{2}\right)
\end{eqnarray}
Because neutrino mass does not involve any hadrons, it has $\Delta B=0$ and therefore, parity violation (i.e. $\Delta I_{3R}\neq 0$) implies that $\Delta L\neq 0$ i.e. the neutrino is a Majorana particle and its small mass is connected to largeness of the parity breaking scale (or the smallness of the strength of $V+A$ currents in weak interaction). A detailed implementation of seesaw and its connection to left-right symmetry breaking can be seen as follows: Suppose that the full gauge group is broken down to the SM group by a Higgs field belonging to $\Delta_R(1,3,2)\oplus  \Delta_L(3,1,2)$ with vev $<\Delta^0_R>=v_R$, then the Yukawa coupling $f\psi_R\psi_R\Delta_R$ leads to a Majorana mass term for the right handed neutrino of magnitude $fv_R$. The Dirac masses arise once the SM gauge group is broken as in GWS model leading to the seesaw formula. This immediately makes it clear how intimately the smallness of neutrino mass in this framework is connected to the parity breaking as well as B-L scale. This makes left-right models a compelling platform for discussing neutrino masses.

Again, to connect with the main message of this article, neutrino mass enhances the case for $B-L$ being the next symmetry of nature and hence possibly the left-right symmetric nature of weak interaction at higher energies. There is active search for the right handed $W_R$'s at LHC via the $\ell\ell jj$ mode\cite{keung}.

It must be emphasized that until direct experimental evidence for left-right symmetric theories such as the signal for a right handed $W_R$ and a heavy right handed neutrino is found at the LHC or alternatively a $Z'$ boson coupled to the $B-L$ current is discovered, the possibility remains open that $B-L$ is not a local but a global symmetry of nature whose breaking could still be at the heart of neutrino masses. In this case, however, there must be a massless Nambu-Goldstone boson present in nature. This particle, called the ``majoron"~\cite{CMP} in literature, can manifest itself in neutrino-less double beta decay process and is being searched for in various experiments. Another signal could be new invisible decay modes of the 125 GeV Higgs boson.

\section{Neutron-antineutron oscillation, Majorana neutrino connection } It is clear from equation \ref{BL},  that parity violation ($\Delta I_{3R}\neq 0$) can also lead to baryon number violation since B is part of the electric charge formula. In fact if $\Delta I_{3R}=1$ which is true if the Higgs field that breaks parity is an $SU(2)_R$ triplet with $B-L=2$ as in the above derivation of seesaw formula for neutrinos, then in principle this theory could lead to $\Delta B=2$ baryon number violating process. There are several such processes e.g., $pp\to K^+K^+, \pi^+\pi^+$, as well as  neutron-antineutron oscillation. The last process is quite interesting as it implies that neutrons traveling in free space can spontaneously convert to antineutrons and current bounds on the oscillation time for this process is $\tau_{n\bar{n}}\geq .8\times 10^8$ sec. It is interesting that even though the oscillation time is about 2 years, all nuclei are stable due to a potential energy difference between neutron and anti-neutron in the nucleus. For a discussion of this and other issues related to neutron-anti-neutron oscillation, see the review~\cite{rabi}. There is now a plan to redo the search for this process at a higher level of sensitivity~\cite{nnbarexpt}.

The question that has to be tackled is whether there exist a theory that combines neutrino mass via the seesaw mechanism which predicts an observable $\tau_{n\bar{n}}$ and yet keeps the proton is stable. One example of such a theory was presented in 1980 in~\cite{MM1}. This model presents an embedding of the left-right seesaw model into a quark lepton unified framework using the gauge group~\cite{PS} $SU(2)_L\times SU(2)_R \times SU(4)_c$  with the symmetry breaking suggested in Ref. 10 rather than in Ref.13. The unified quarks-lepton multiplet is given by $\Psi_L(2,1,4)\oplus \Psi_R(1,2,4)$ with $\Psi$ given by
\begin{eqnarray}
\Psi~=~\left(\begin{array}{cccc} u_1 & u_2 & u_3 & \nu_e\\d_1 & d_2 & d_3 & e\end{array}\right)
\end{eqnarray}
where subscripts $(1,2,3)$ denote the color index. The sixteen chiral fermions of the $SU(2)_L\times SU(2)_R \times SU(4)_c$ model fit into the sixteen dimensional spinor representation of the SO(10) group~\cite{GFM} which can be the final grand unification group for left-right symmetry as well as $B-L$ gauge symmetry. The symmetry breaking from the group $SU(2)_L\times SU(2)_R \times SU(4)_c$ down to the SM group is achieved by the Higgs fields $\Delta (1,3,10)$ which is the $SU(4)_c$ generalization of the seesaw generating Higgs field $\Delta_R(1,3,+2)$ discussed in the previous section. Without getting into too much group theory details, one can see that the $\Delta_R$ fields must have their quark partners present inside them (denote them by $\Delta_{qq}$), which couple to two quarks. Combined with $\Delta_{\nu_R\nu_R}$ vev breaking B-L symmetry to give seesaw structure for neutrino masses, this leads to the six quark operator $u_Rd_Rd_Ru_Rd_Rd_R$ via the diagram in Fig. 3 below to lead to non-vanishing neutron oscillation amplitude. For multi-TeV scale seesaw, the strength of this operator is of order $G_{\Delta B=2}\sim \frac{\lambda f^3 v_R}{M^6_{\Delta_{qq}}}\sim 10^{-28}$ GeV$^{-5}$ for $f\sim \lambda 10^{-2}$ and $v_R\sim M_{\Delta}\sim 10$ TeV. Once this operator is hadronically dressed, it gives $\tau_{n\bar{n}}\sim 10^{8-10}$ sec., which is in the observable range with currently available neutron sources around the world.

\begin{figure}
\centerline{\includegraphics[width=4.5cm]{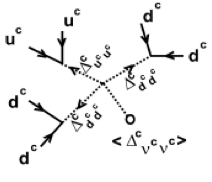}}\centerline{\includegraphics[width=6.5cm]{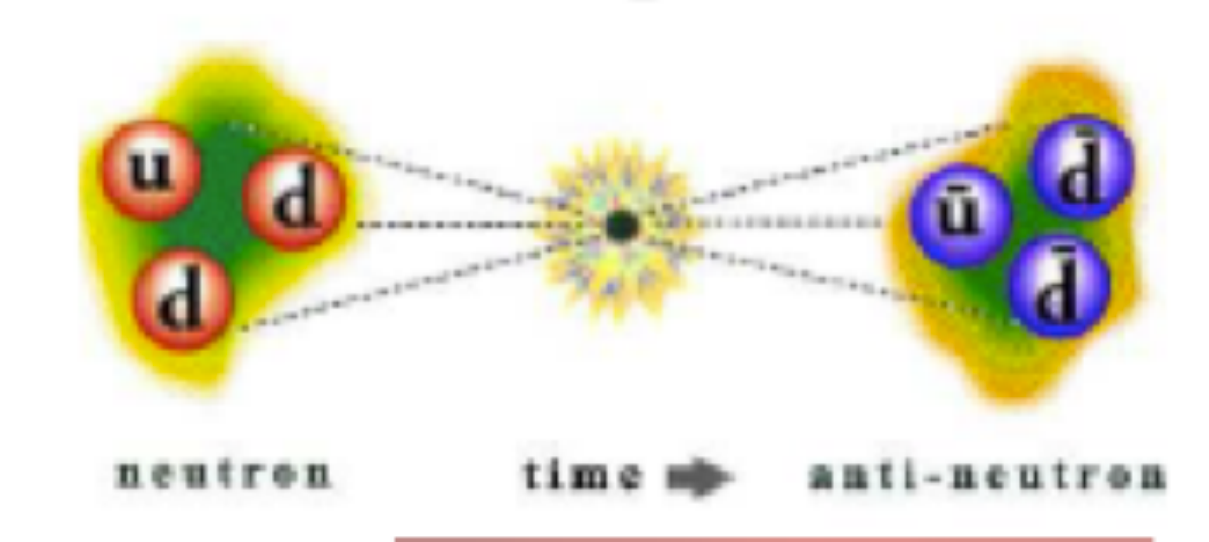}}
\caption{ Feynman diagram for neutron oscillation and Neutrino mass $n\bar{n}$ connection in a quark-lepton unified left-right seesaw model} \label{ra_fig1}
\end{figure}

\section{Role of $B-L$ in baryon number violation} We saw above two extensions of standard model which lead to two specific examples of B-L violation: one for neutrino mass and another for neutron-anti-neutron oscillation. One can ask the question as to whether we can say in a model independent way about the B-L violation in SM extensions to high scale without detailed specification. One way to explore this would be to consider higher dimensional B and L violating operators that are invariant under the SM gauge group. This discussion was carried out many years ago by Weinberg~\cite{W1} and Wilczek and Zee~\cite{W2}. They pointed out that there is one d+5 operator invariant under the SM involving only the SM fields i.e. ${\cal O}_1= LHLH$, which leads to the Majorana mass for neutrinos once electroweak symmetry is broken down by the Higgs vev and changes L by two units. The Left-right model is a ultra-violet complete model that leads to this operator below the right handed scale. At the level of $d=6$, there are five operators: $ {\cal O}_2=QQQL$, ${\cal O}_3=Q^TC^{-1}\vec{\tau}Q\cdot Q^TC^{-1}\vec{\tau}L$, ${\cal O}_4= QL(u^cd^c)^*$ and ${\cal O}_5=QQ(u^ce^c)^*$, ${\cal O}_6=u^cu^cd^ce^c$. The interesting property is that they all conserve the B-L quantum number and proton decay of type $p\to e^+\pi^0$. It is interesting that operators at the level of $d=7$ break baryon number in such a way that they lead to B-L=2 proton decay models e.g. $n\to e^-\pi^+$~\cite{d7,baburabi}. 
\begin{eqnarray}
{\cal {O}}'_1 &=& (d^c u^c)^* (d^c L_i)^* H^*_j \epsilon_{ij},~~~~~~{\cal {O}}'_2 = (d^c d^c)^* (u^c L_i)^* H^*_j \epsilon_{ij}, \nonumber \\
{\cal {O}}'_3 &=& (Q_i Q_j)(d^c L_k)^*H^*_l \epsilon_{ij} \epsilon_{kl},~~
{\cal {O}}'_4 = (Q_i Q_j) (d^c L_k)^*H^*_l (\vec{\tau} \epsilon)_{ij}\cdot (\vec{\tau}\epsilon)_{kl}, \nonumber \\
{\cal {O}}'_5 &=& (Q_i e^c) (d^c d^c)^*H^*_i,~~~~~~~~~~
{\cal {O}}'_6 = (d^c d^c)^*(d^c L_i)^* H_i, \nonumber \\
{\cal {O}}'_7 &=& (d^c D_\mu d^c)^*(\overline{L}_i \gamma^\mu Q_i),~~~~~~{\cal {O}}'_8 = (d^c D_\mu L_i)^*(\overline{d^c} \gamma^\mu Q_i), \nonumber \\
{\cal {O}}'_9 &=& (d^c D_\mu d^c)^* (\overline{d^c} \gamma^\mu e^c)~.
\label{dim7}
\end{eqnarray}

A well known example of a UV complete theory where the above $d=6$ operators emerge is the minimal $SU(5)$ model~\cite{gg}, where after symmetry breaking to the standard model,  $B-L$ remains a good symmetry. Examples of UV complete theories where the above $d=7$ as well as the $d=6$ operators arise have also been recently discussed~\cite{baburabi} based on SO(10) grand unified theories with the seesaw mechanism. These models lead to both $B-L$ conserving and violating nucleon decay.

\section{$B-L$ breaking and how the presence of nucleon decay and $n\bar{n}$ together imply Majorana neutrinos}
A key question in physics beyond the standard model (BSM) is whether neutrinos are Dirac or Majorana type fermions. The answer to this question will dictate the path of BSM physics. The most direct experimental way is to settle this question is by searching for neutrino-less double beta decay($\beta\beta_{o\nu}$)  of  certain nuclei. Rightly therefore, there is intense activity in this field at the moment at various laboratories around the world. The current round of experiments, however,  are sensitive enough to probe only a small region of neutrino masses and that too provided the neutrino mass ordering is inverted type~\cite{pascoli}. Even a large bulk of the inverted mass hierarchy region cannot be reached by the current experiments. On the other hand, if neutrino mass hierarchy is normal, it is indeed very unlikely that we will know the answer to this very important question from searches for neutrinoless double beta decay for a long time. It is therefore not without interest to search for alternative experimental strategies to answer this question. Below we suggest that one such strategy is to put renewed effort on the search for baryon number violation~\cite{BM1}.

We saw above that $B-L$ breaking connects  Majorana neutrinos and baryon number violation via the electric charge formula if it is a local symmetry. Pursuing this line of thinking, it should be possible to use only B-violating processes, to experimentally resolve the above key questions of the nature of neutrino masses. As suggested recently~\cite{BM1},  consider two different  B-violating processes such as  $p\to e^+\pi^0$ and neutron oscillation or two $B$-violating nucleon decay modes, one of which conserves $B-L$ and another that breaks it such as $p\to e^+\pi^0$ (which is of $B-L=0$ type) and $n\to e^-\pi^+$ (which has $B-L=2$)~\cite{BM1}. Simultaneous discovery of any pair of these processes  will imply that there must be neutrino-less double beta decay  via a typical Feynman diagram of type in Fig. 4 and once neutrino-less double beta decay has a nonzero amplitude (whether obtained directly or indirectly), no matter how small it is, it will imply a Majorana mass for the neutrino~\cite{SV}. This can provide an alternative way to experimentally answer the Majorana or Dirac nature of the neutrino regardless of how small the effective neutrino mass contributing to $\beta\beta_{0\nu}$ decay is. 

\begin{figure}
\centerline{\includegraphics[width=6.5cm]{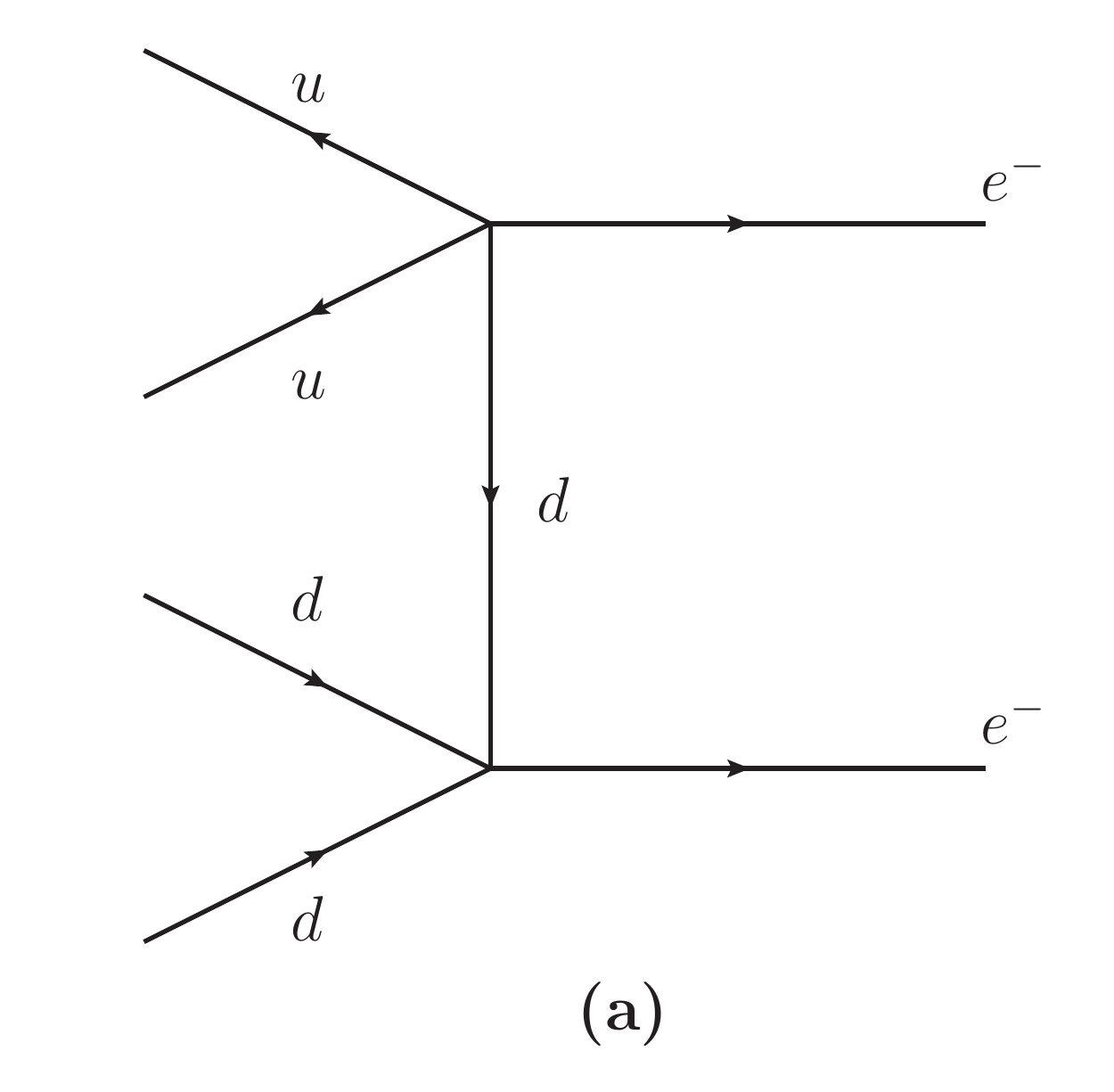}}
\caption{  How discovery of $n\to e^-\pi^+$  and $p\to e^+\pi^0$ implies a nonzero amplitude for neutrinoless double beta decay} \label{ra_fig1}
\end{figure}

Another way  to settle the same issue is to invoke the sphaleron operator of the standard model and combine it with proton decay and $n-\bar{n}$ oscillation. To see how this argument goes,  note that the non=perturbative effects of the standard model lead to B-violation  given by the operator  $QQQQQQQQQLLL$. Here $Q$ and $L$ are the $SU(2)_L$ doublets. We can formally rewrite this operator as product of three operators i.e.$QQQQQQ$,  $QQQL$ and $LL$. Explicitly rewriting this by expanding $Q\equiv (u,d)$ and $L\equiv (\nu_e, e)$ and using quark mixings to change generations, we get for one of the pieces of the sphaleron operator to be $uddudd\cdot uude\cdot \nu\nu$. The strength of this operator is of course highly suppressed. However, as a matter of principle, note that the first part is the piece that contributes to $n\bar{n}$ oscillation, second part to the $p\to e^+\pi^0$ decay and the last part to Majorana mass for the neutrinos. One can represent this in terms a triangle which I call ``B-L triangle'' (see Fig.5). The advantage here is that this combination directly gives Majorana neutrino mass without the intermediary of neutrinoless double beta decay.

\begin{figure}
\centerline{\includegraphics[width=6.5cm]{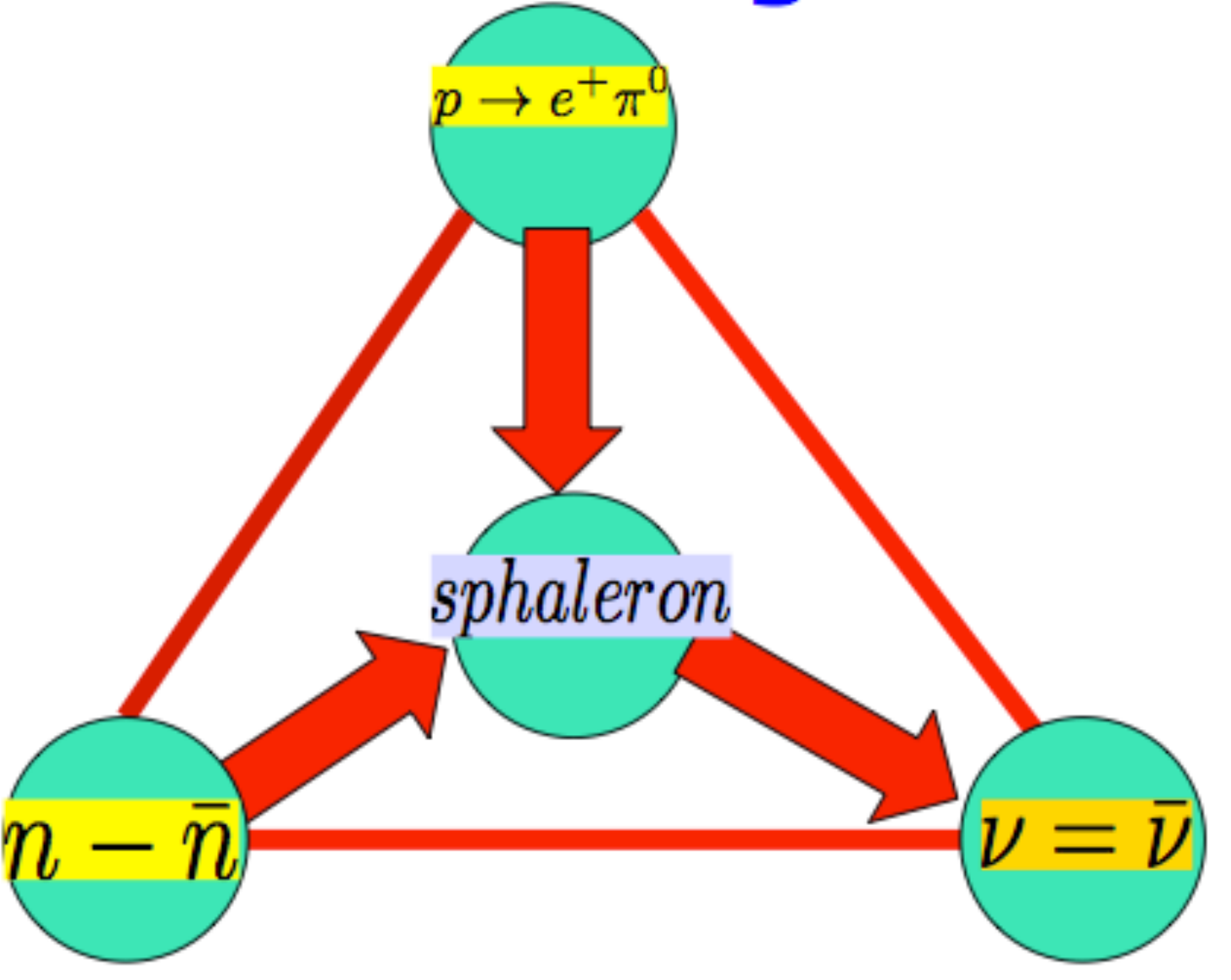}}
\caption{ ``B-L Triangle'' explains how discovering neutron oscillation and proton decay implies neutrinos are Majorana fermions} \label{ra_fig1}
\end{figure}

\section{$B-L$, supersymmetry and neutralino as a dark matter}
A final application of local $B-L$ symmetry concerns an understanding of the widely discussed suggestion that in supersymmetric models, the lightest superpartner (LSP) of the standard model particles may play the role of dark matter of the universe. While this suggestion has led to a great deal of dark matter related activity in both theory and experiment, a key question is that in the minimal supersymmetric standard model (MSSM), there are interactions, the so-called R-parity violating ones, which imply that the LSP is actually unstable. So is there a natural extension of MSSM that would lead to a stable dark matter. It was pointed out in mid-eighties and early 90s that if MSSM is extended to include a local $B-L$ symmetry and if $B-L$ symmetry is broken by a Higgs field that has $B-L=2$, then R-parity ism indeed automatic~\cite{RPC}. A simple way to see this is to realize that R-parity symmetry can be written as: $R=(-1)^{3(B-L)+2S}$ and when R-parity is violated by a Higgs field with $B-L=2$, $R$ remains unbroken. It is interesting that the same Higgs field that gives Majorana mass to the right handed neutrinos also preserves R-parity symmetry leading to a stable neutralino dark matter.

\section{Summary}
In summary, we have provided a broad brush overview of the history of B-L as a new symmetry in particle physics and how in recent years following the discovery of neutrino masses, interest in this possible new symmetry has grown enormously. In particular, its connection to both neutrino mass and baryon number violation have provided new insights into physics beyond the standard model. All these have to be confirmed experimentally. At the same time, there are phenomenological studies of many different aspects of this symmetry.
To  summarize the efforts to unravel the degree of freedom corresponding to local B-L  symmetry experimentally, i mention only a few topics. Deciphering whether neutrinos are Majorana fermions is a direct confirmation of whether B-L symmetry is broken or not. This does not say whether it is a global or local symmetry. Furthermore, by itself,  discovery of $\beta\beta_{0\nu}$ decay cannot tell where the scale of B-L symmetry breaking is. Supplemented by a discovery (or non-discovery) of neutron oscillation, one can get an idea about a possible range (or exclude a possible range) of this scale but not the actual scale. The most definitive way to discover the scale of B-L symmetry is to directly search for the gauge boson associated with this in the collider such as the LHC~\cite{BLexpt}. The same could also be inferred from a discovery of the $W_R$ combined with a Majorana right handed neutrino. Such searches are currently under way at the LHC ~\cite{LHC}.

\section*{Acknowledgement} This work is supported by the National Science Foundation grant No. PHY-1315155. I am grateful to Tom Ferbel for a careful reading of the manuscript.


\newpage

\begin{thebibliography}{99}


\bibitem{gz} M.~Gell-Mann,
  Phys.\ Lett.\  {\bf 8}, 214 (1964);  G.~Zweig,
 CERN Geneva - TH. 401 (REC.JAN. 64) 24p
 

\bibitem{gmo} A. Gamba, R. E. Marshak, and S. Okubo, Proc Natl Acad Sci (U S A.) 45: 881Ð885 (1959).

\bibitem{REM} For a review and history of these ideas, see R. E. Marshak, Prog. Theor. Phys. {\bf 85}, 61 (1985).

 \bibitem{gws} S.~L.~Glashow,
  Nucl.\ Phys.\  {\bf 22}, 579 (1961);  S.~Weinberg,
  Phys.\ Rev.\ Lett.\  {\bf 19}, 1264 (1967); A. Salam, Proceedings of the Nobel symposium, ed. N. Svartholm et al (1968).
  
   \bibitem{bg}  J.~D.~Bjorken and S.~L.~Glashow,
  Phys.\ Lett.\  {\bf 11}, 255 (1964).
  
  \bibitem{BEHGHK}  F. Englert and R. Brout, Phys. Rev. Lett. {\bf 13}, 321 (1964);
 Peter W. Higgs, Phys. Rev. Lett. {\bf 13}, 508 (1964); G. S. Guralnik, C.R. Hagen and T.W.B. Kibble,  Phys. Rev. Lett. {\bf 13}, 585 (1964).
  
  
  \bibitem{GIM} S.~L.~Glashow, J.~Iliopoulos and L.~Maiani,
  Phys.\ Rev.\ D {\bf 2}, 1285 (1970).
  
  
  \bibitem{adler} S. L. Adler, Phys. Rev. {\bf 177},  2426 (1969);  J. S. Bell and R. Jackiw, Nuovo Cimento {\bf A 60},  47 (1969); S. L. Adler and W. A. Bardeen, Phys. Rev. {\bf 182} 1517 (1969); W. Bardeen,  Phys. Rev. {\bf 184}  1848 (1969).
  
  \bibitem{th} G.~'t Hooft,
  Phys.\ Rev.\ Lett.\  {\bf 37}, 8 (1976).
  
  
 
  \bibitem{mp}   J. C. Pati and A. Salam, Phys. Rev. {\bf D 10}, 425 (1974); R. N. Mohapatra and J. C. Pati, Phys. Rev.{\bf D 11}, 566, 2558 (1975); G. Senjanovi\'c and R. N. Mohapatra, Phys. Rev. {\bf D 12}, 1502 (1975).
  
  \bibitem{MM} R.~E.~Marshak and R.~N.~Mohapatra,
  Phys.\ Lett.\ B {\bf 91}, 222 (1980).
  
  \bibitem{david} A.~Davidson,
  Phys.\ Rev.\ D {\bf 20}, 776 (1979).
  
   \bibitem{seesaw} P. Minkowski, Phys. Lett. {\bf  B 77},  421 (1977);  M. Gell-Mann, P. Ramond and R. Slansky, in Supergravity, eds. D. Freedman et al.
(North-Holland, Amsterdam, 1980); T. Yanagida, in proc. KEK workshop, 1979 (unpublished); S. L. Glashow, Cargese lectures, (1979); R.N. Mohapatra and G. Senjanovi\'c, Phys. Rev. Lett. {\bf 44}, 912 (1980).

\bibitem{majorana} For recent pedagogical reviews of the developments in neutrino physics, see L. Maiani,   arXiv:1406.5503, ``{\it  Notes from the Ettore Majorana Lectures}"; 
S. M. Bilenky, arXiv:1408.1432 [hep-ph] .

\bibitem{keung}  W.~Y.~Keung and G.~Senjanovic,
  Phys.\ Rev.\ Lett.\  {\bf 50}, 1427 (1983).
 

\bibitem{CMP} Y.~Chikashige, R.~N.~Mohapatra and R.~D.~Peccei,
  Phys.\ Lett.\ B {\bf 98}, 265 (1981).
  
  
  \bibitem{MM1} R.~N.~Mohapatra and R.~E.~Marshak,
  Phys.\ Rev.\ Lett.\  {\bf 44}, 1316 (1980)
  [Erratum-ibid.\  {\bf 44}, 1643 (1980)].
 
  
 \bibitem{rabi}  R.~N.~Mohapatra,
  J.\ Phys.\ G {\bf 36}, 104006 (2009)
  [arXiv:0902.0834 [hep-ph]].
  
  \bibitem{nnbarexpt} K.~Babu, S.~Banerjee, D.~V.~Baxter, Z.~Berezhiani, M.~Bergevin, S.~Bhattacharya, S.~Brice and T.~W.~Burgess {\it et al.},
  arXiv:1310.8593 [hep-ex].
 
  
  \bibitem{PS} J. C. Pati and A. Salam, ref. 8.
  
  \bibitem{GFM}  H.~Fritzsch and P.~Minkowski,
  Annals Phys.\  {\bf 93}, 193 (1975);
H. Georgi, in {\it Particles and Fields}, ed. C. E. Carlson, A. I. P. (1975);  
  
  \bibitem{W1} S.~Weinberg,
  Phys.\ Rev.\ Lett.\  {\bf 43}, 1566-1570 (1979).
  
  
  \bibitem{W2}  F.~Wilczek, A.~Zee,
  Phys.\ Rev.\ Lett.\  {\bf 43}, 1571-1573 (1979).
  
  \bibitem{d7} H.~A.~Weldon and A.~Zee,
  Nucl.\ Phys.\ B {\bf 173}, 269 (1980);  R.~E.~Marshak and R.~N.~Mohapatra,
  In *Coral Gables 1980, Proceedings, Recent Developments In High-energy Physics*, 277-287 and Virginia Polytech.Blacksburg - VPI-HEP-80-02 (80,REC.FEB) 15 P. (002309);
K.~S.~Babu and R.~N.~Mohapatra,
  Phys.\ Rev.\ Lett.\  {\bf 109}, 091803 (2012); S.~M.~Barr and X.~Calmet,
  Phys.\ Rev.\ D {\bf 86}, 116010 (2012).
  
  \bibitem{gg} H.~Georgi and S.~L.~Glashow,
  Phys.\ Rev.\ Lett.\  {\bf 32}, 438 (1974).
  
  \bibitem{baburabi} K.~S.~Babu and R.~N.~Mohapatra,
  Phys.\ Rev.\ Lett.\  {\bf 109}, 091803 (2012).
  
  \bibitem{pascoli} S.~M.~Bilenky, S.~Pascoli and S.~T.~Petcov,
  Phys.\ Rev.\ D {\bf 64}, 053010 (2001); F.~Feruglio, A.~Strumia and F.~Vissani,
  Nucl.\ Phys.\ B {\bf 637}, 345 (2002)
  [Addendum-ibid.\ B {\bf 659}, 359 (2003)]
 
  

 \bibitem{BM1} K.~S.~Babu and R.~N.~Mohapatra,
  arXiv:1408.0803 [hep-ph].
 
 \bibitem{SV} J.~Schechter and J.~W.~F.~Valle,
  Phys.\ Rev.\ D {\bf 25}, 2951 (1982).
  
  
  \bibitem{BLexpt}  For a small sample of recent papers on B-L gauge boson searches, W.~Emam and S.~Khalil,
  Eur.\ Phys.\ J.\ C {\bf 52}, 625 (2007)
  [arXiv:0704.1395 [hep-ph]]; L.~Basso,
  arXiv:1106.4462 [hep-ph]; L.~Basso, A.~Belyaev, S.~Moretti and G.~M.~Pruna,
  Nuovo Cim.\ C {\bf 33N2}, 171 (2010)
  [arXiv:1002.1214 [hep-ph]]; K.~Huitu, S.~Khalil, H.~Okada and S.~K.~Rai,
  Phys.\ Rev.\ Lett.\  {\bf 101}, 181802 (2008)
  [arXiv:0803.2799 [hep-ph]]; for a review of general $Z^\prime$ searches, see, P.~Langacker,
  Rev.\ Mod.\ Phys.\  {\bf 81}, 1199 (2009).
  
  \bibitem{LHC} V. Khachatryan et al. [CMS Collaboration], arXiv:1407.3683 [hep-ex]; G.~Aad {\it et al.}  [ATLAS Collaboration],
  Eur.\ Phys.\ J.\ C {\bf 72}, 2056 (2012)
  [arXiv:1203.5420 [hep-ex]].
 
\bibitem{RPC}  R.~N.~Mohapatra,
  Phys.\ Rev.\ D {\bf 34}, 3457 (1986); A.~Font, L.~E.~Ibanez and F.~Quevedo,
  Phys.\ Lett.\ B {\bf 228}, 79 (1989);
 S.~P.~Martin,
  Phys.\ Rev.\ D {\bf 46}, 2769 (1992); C.~S.~Aulakh, A.~Melfo, A.~Rasin and G.~Senjanovic,
  Phys.\ Lett.\ B {\bf 459}, 557 (1999).
 
  





\end{thebibliography}
\end{document}

To see how this argument goes, note that note that the sphaleron operator is given by $QQQQQQQQQLLL$. Here $Q$ and $L$ are the $SU(2)_L$ doublets. We can rewrite this as product of three parts i.e.$QQQQQQ$,  $QQQL$ and $LL$. Explicitly rewriting this by expanding $Q\equiv (u,d)$ and $L\equiv (\nu_e, e)$, we get for one of the pieces to be $uddudd\cdot uude\cdot \nu\nu$. Note that the first part is the piece that contributes to $n\bar{n}$ oscillation, second part to the $p\to e^+\pi^0$ decay and the last part to Majorana mass for the neutrinos. One can represent this in terms a triangle which I call ``B-L triangle'' (see Fig.2).